\documentclass[fleqn,10pt]{wlscirep}
\title{Cascading failures in coupled networks with both inner-dependency and inter-dependency links}

\author[1,*]{Run-Ran Liu}
\author[2,3]{Ming Li}
\author[1]{Chun-Xiao Jia}
\author[2]{Bing-Hong Wang}
\affil[1]{Alibaba Research Center for Complexity Sciences, Hangzhou Normal University, Hangzhou, 311121, People's Republic of China}
\affil[2]{Department of Modern Physics, University of Science and Technology of China, Hefei, 230026, People's Republic of China}
\affil[3]{Department of Applied Physics, Hong Kong Polytechnic University, Hung Hom, Hong Kong}

\affil[*]{runranliu@163.com}

\begin{abstract}
We study the percolation in coupled networks with both inner-dependency and inter-dependency links, where the inner- and inter-dependency links represent the dependencies between nodes in the same or different networks, respectively. We find that when most of dependency links are inner- or inter-ones, the coupled networks system is fragile and makes a discontinuous percolation transition. However, when the numbers of two types of dependency links are close to each other, the system is robust and makes a continuous percolation transition. This indicates that the high density of dependency links could not always lead to a discontinuous percolation transition as the previous studies. More interestingly, although the robustness of the system can be optimized by adjusting the ratio of the two types of dependency links, there exists a critical average degree of the networks for coupled random networks, below which the crossover of the two types of percolation transitions disappears, and the system will always demonstrate a discontinuous percolation transition. We also develop an approach to analyze this model, which is agreement with the simulation results well.
\end{abstract}
\begin{document}

\flushbottom
\maketitle

\thispagestyle{empty}

\section*{Introduction}

In the past decade, the robustness of isolated networks has been extensively studied \cite{nw3,nw8,nw9}. Recently, based on the motivation that many real-world complex systems, such as physical, social, biological, and infrastructure systems, are becoming significantly more dependent on each other, the robustness of coupled networks has been studied by means of percolation in interdependent networks \cite{idn1}. In these works, the inter-dependency links have been proposed to represent the dependencies of nodes between different networks. Consequently, the failure of a node will result in the failure of the node connected to it by a dependency link. It has been recognized that the inter-dependency makes the coupled system more fragility than a single network \cite{idn1,vespignani2010}, especially for the system with multiple networks coupled together \cite{Gao2011a,Gao2011b,non1,non2}, and demonstrates a discontinuous percolation transition.

Along this pioneering work, interdependent networks with different topological properties, coupling method and attack strategies have been studied extensively in the past few years, such as partially dependency \cite{idn2}, inter-similarity \cite{idn3,idn4}, multiple support-dependency relations \cite{idn5}, targeted attack \cite{idn6} and localized attack\cite{loc1,loc2}, assortativity \cite{idn8,Valdez,assor}, clustering \cite{cluster1,cluster2}, degree distribution \cite{de1,de2}, and spatially embedded networks\cite{idn9,Bashan2013,Shekhtman,Danziger}. All these works further demonstrate that the fragility of the networks when they are dependent on each other.

On the other hand, to reflect the strongly dependency of units inside a system, percolation in networks with inner-dependency links has also attracted a great attention \cite{sdl1,lim1}. Similar with the interdependent networks, the iterative process of cascading failures caused by connectivity and dependency links will also lead to a discontinuous percolation transition, rather than the well-known continuous phase transition in isolated networks, which has a devastating effect on the network stability. Furthermore, with a view to that more than two nodes depend on each other, dependency group is often used to replace the dependency link in the study of percolation in isolated networks with dependency \cite{sdl2,sdl3,lim2,lim3}.

However, the previous studies of the percolation in networks with dependency are all based on the assumption that the networks contain either inner-dependency links or inter-dependency links \cite{rvn}. For a real network system, some nodes may depend on nodes outside the networks, and some inside. That is to say that the inner- and inter-dependency links could exists in a coupled networks system simultaneously. For example, in a trading network, some companies may depend on each other due to supply and demand balance. On the other hand, some companies could depend on some units in a financial network, which forms by banks, investors, and so on. Although the effects of the two types of dependencies on the network stability have been explored separately, there is still lack of unified understanding of various robustness properties of the coupled networks due to the coaction of the two types of links. In this paper, we will develop a model to study the robustness of such networks, i.e., networks with both inner- and inter-dependency links.

This paper is organized as follows. In the next section, we will give the model and general formalism using generating function techniques. After that, we will give study our model on coupled random networks system and coupled scale-free networks as examples. At the same time, the simulation results will be presented to test the analysis results. In the last section, we will summary our findings in this paper.

\section*{Results}
\subsection*{Model and general formalism}
We consider two coupled networks $A$ and $B$ with degree distributions $p_{k}^{A}$ and $p_{k}^{B}$, respectively, and each node has exactly one dependency link (inner- or inter- dependency link), where the dependency link means that the two nodes connected by it depend on each other, one of which fails, the other will fail too. Assuming that the two networks have the same size $N$, there are $N$ dependency links in the network system. Specifically, a fraction $\beta$ of the dependency links are set as the inter-dependency links, others are the inner-dependency links. For inter-dependency links, the two stubs (nodes) are chosen randomly in the two networks, respectively, and in the same networks for inner-dependency links. When $\beta \rightarrow 0$, there is no dependency between the two networks and the model will reduce to the model of the single network with dependency link density $q=1$ in ref.\cite{sdl1}. When $\beta \rightarrow 1$, our model will reduce to the original model of interdependent network proposed in ref.\cite{idn1}.

We want to study the robustness of such coupled system after an initial attack of a fraction, $1-p$, of nodes in network $A$. The failure of a node in network $A$ will lead to the failure of its dependency partner no matter it is in network $A$ or network $B$, even though it still connects to the network by connectivity links. The failures of nodes in network $B$ have the similar consequence. On the other hand, the failures of nodes or their connectivity links may also cause the other nodes to disconnect from the networks, which is also considered as failure. Therefore, after the initial attack in network $A$, the two cascading processes (dependency and connectivity) will occur alternately in networks $A$ and $B$ until no further splitting and node removal can occur.

Here, we focus on the size of the giant component of the two networks, $S^A$ and $S^B$, which are the probability that a randomly chosen node belongs to the giant component of the final network $A$ or $B$, respectively. Note that $S^A$ is generally different from $S^B$ due to the initial node removal. To solve this model as the method used in refs.\cite{fs1,fs2}, we need two auxiliary parameters $R^A$ and $R^B$, which give the probability that the node, arriving at by following a randomly chosen link in network $A$ or $B$, belongs to the giant component of the final network $A$ or $B$. Then, in the steady state, $S^A$ satisfies
\begin{equation}
S^{A} = p^{2}(1-\beta) (f^{A})^{2}+ p \beta f^{A}f^{B}. \label{Sa}
\end{equation}
Here, $f^{A}=1-G_{0}^{A}(1-R^{A})$ and $f^{B}=1-G_{0}^{B}(1-R^{B})$ with $G_{0}^{A}(x)=\sum_kp_{k}^{A}x^k$ and $G_{0}^{B}(x)=\sum_kp_{k}^{B}x^k$ denoting the corresponding generating functions of the degree distributions of networks $A$ and $B$, respectively. Obviously, $f^{A}$ ($f^{B}$) means the probability that a randomly chosen node in network $A$ ($B$) belongs to the giant component of network $A$ ($B$)\cite{nm}. Since the two stubs of a dependency link are chosen randomly, $(f^{A})^2$ and $f^{A}f^{B}$ express that a node in network $A$ and its dependency partner in network $A$ or $B$ (with a fraction $\beta$ or $1-\beta$) belongs to the giant component, simultaneously. In addition, $p^2$ expresses that the node and its dependency partner in network $A$ are preserved after the initial removal.

Similarly, $S^{B}$ can be written as
\begin{equation}
S^{B} = (1-\beta) (f^{B})^{2}+ p \beta f^{A}f^{B}. \label{Sb}
\end{equation}
Since the initial attack only takes place in network $A$, the first term of the right side of eq.(\ref{Sb}) is different with that of eq.(\ref{Sa}).

To solve eqs.(\ref{Sa}) and (\ref{Sb}), we need the equations for $R^A$ and $R^B$, which can be obtained by considering the branch process in the two networks \cite{nm},
\begin{eqnarray}
R^A &=&  p^{2} (1-\beta)[1-G_{1}^A(1-R^A)][1-G_{0}^A(1-R^A)] + p \beta [1-G_{1}^A(1-R^A)][1-G_{0}^B(1-R^B)], \\ \label{Ra}
R^B &=& (1-\beta)[1-G_{1}^B(1-R^B)][1-G_{0}^B(1-R^B)]+ p \beta [1-G_{1}^B(1-R^B)][1-G_{0}^A(1-R^A)], \label{Rb}
\end{eqnarray}
where $G_{1}^A(x)=\sum_{k}p_{k}^{A}kx^{k-1}/\langle k\rangle^A =G^{A\prime}_{0}(x)/G^{A\prime}_{0}(1)$ is the corresponding generating function of the underlying branching processes of network $A$, and the brackets $\langle \cdots \rangle$ denote an average over the degree distribution $p_{k}^{A}$. Similarly, $G_{1}^B(x)=\sum_{k}p_{k}^{B}kx^{k-1}/\langle k\rangle^B =G^{B\prime}_{0}(x)/G^{B\prime}_{0}(1)$. Given arbitrary degree distributions $p_{k}^A$, $p_{k}^B$ and the fraction of initial removal $1-p$, we can solve eqs. (\ref{Sa})-(\ref{Rb}) to obtain the order parameters $S^A$ and $S^B$.

\subsection*{Random networks}
Next, we will study two coupled random networks with the same Poisson degree distribution $p_k=\frac{e^{-\langle k \rangle}\langle k \rangle ^k}{k!}$ in details \cite{tmgt}, where $\langle k \rangle$ is the average degree. In this case, the generating functions of the two networks take a simple form $G_{0}^A(x)=G_{1}^A(x)=G_{0}^B(x)=G_{1}^B(x)=e^{-\langle k \rangle (1-x)}$. Therefore, we have $R^{A}=S^{A}$ and $R^{B}=S^{B}$. This yields
\begin{eqnarray}
S^A &= &p^{2} (1-\beta)(1-e^{-\langle k \rangle S^A})^{2} + p \beta (1-e^{-\langle k \rangle S^A})(1-e^{-\langle k \rangle S^B}), \label{Saer} \\
S^B &= &(1-\beta)(1-e^{-\langle k \rangle S^B})^{2} + p\beta (1-e^{-\langle k \rangle S^A})(1-e^{-\langle k \rangle S^B}). \label{Sber}
\end{eqnarray}
For $\beta=0$, one obtains $S^A= p^2 (1-e^{-\langle k\rangle S^A})^2$ and $S^B= (1-e^{-\langle k\rangle S^B})^2$. This covers the equations found in refs. \cite{sdl2,sdl3}. In this case, the percolation transition of network $A$ is discontinuous, and network $B$ has nothing to do with the fraction of initial preserved nodes $p$. For another case $\beta=1$, one can also find that $S^A=S^B=p (1-e^{-\langle k\rangle S^A})^2$, which coincides with the result of the interdependent networks\cite{idn1}.

\begin{figure}
\scalebox{0.6}[0.6]{\includegraphics{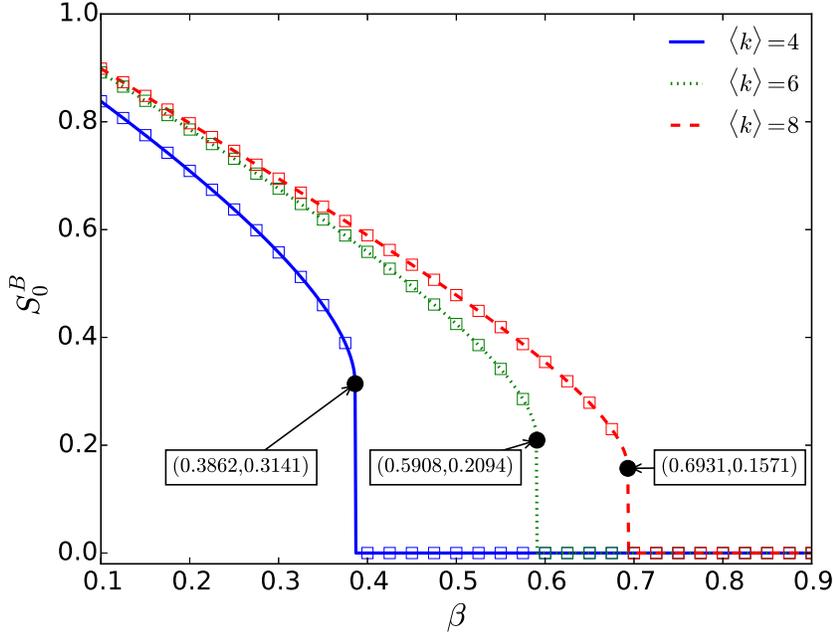}} \caption{ (Colour online) The minimum values of $S^{B}$, labeled as $S^{B}_{0}$, as a function of the parameter $\beta$ for different average degrees. The value of $S^{B}_{0}$ jumps from $S^{B}_{0}\approx1.2564/\langle k\rangle$ to zero abruptly at the critical point $\beta_{c}'\approx1-2.4554/\langle k\rangle$. The lines denote the numerical solutions and the symbols denote the simulation results from $20$ time realizations on networks with $10^5$ nodes.} \label{f1}
\end{figure}

Next, we discuss the solution of eqs.(\ref{Saer}) and (\ref{Sber}) to obtain the percolation properties of this system. In general, eqs. (\ref{Saer}) and (\ref{Sber}) have a trivial solution at $(S^{A}=0,S^{B}=0)$, which means that the two networks $A$ and $B$ are completely fragmented. In addition, there is another trivial solution $(S^{A}=0,S^{B}>0)$ for eqs.(\ref{Saer}) and (\ref{Sber}) as the initial node removal is only for network $A$. Let $S^{A}=0$ in eq.(\ref{Sber}), we can get the trivial solution of $S^{B}$,
\begin{equation}
S^{B}_{0}=(1-\beta)(1-e^{-\langle k \rangle S^{B}_{0}})^{2}. \label{sb0}
\end{equation}
Here, we use $S^{B}_{0}$ instead of $S^{B}$ to avoid confusion. As the numerical solution of eq.(\ref{sb0}) shown in Fig.\ref{f1}, above a critical point $\beta_{c}'\approx1-2.4554/\langle k\rangle$, the minimum values $S^{B}_{0}=0$, which is equivalent to the trivial solution $(S^{A}=0,S^{B}=0)$, and means network $B$ is completely fragmented with the fragmented of network $A$. And below the critical point $\beta_{c}'$, $S^{B}_{0}>0$, which means that network $B$ is still functioning, although network $A$ is completely fragmented.

\begin{figure}
\scalebox{0.6}[0.6]{\includegraphics{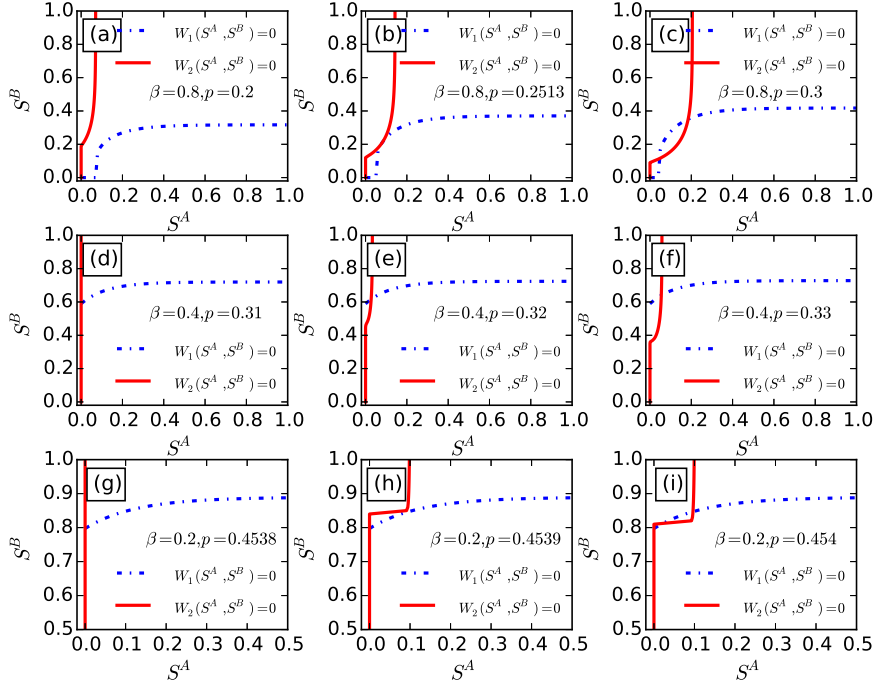}} \caption{ (Colour online) Graphical solutions for eqs.(\ref{Saer}) and (\ref{Sber}) with $\langle k\rangle=8$. (a)-(c), $\beta=0.8>\beta_c'$, $p_c\approx 0.2513$ with nonzero $S^A_c$ and $S^B_c$. (d)-(f), $\beta=0.4<\beta_c'$, $p_c\approx 0.3136$ with $S^A_c=0$ and nonzero $S^B_c$. (g)-(i), $\beta=0.2<\beta_c'$. $p_c\approx 0.4539$ with nonzero $S^A_c$ and $S^B_c$. } \label{f2}
\end{figure}

In order to discuss the nontrivial solutions, we construct two functions based on eqs. (\ref{Saer}) and (\ref{Sber}),
\begin{eqnarray}
W_{1}(S^A,S^B) &= &S^A -  p^{2} (1-\beta)(1-e^{-\langle k \rangle S^A})^{2} - p \beta (1-e^{-\langle k \rangle S^A})(1-e^{-\langle k \rangle S^B}),\\
W_{2}(S^A,S^B) &= &S^B - (1-\beta)(1-e^{-\langle k \rangle S^B})^{2} - p \beta (1-e^{-\langle k \rangle S^A})(1-e^{-\langle k \rangle S^B}).
\end{eqnarray}
The nontrivial solution of $S^A$ and $S^B$ can be presented by the crossing points of the cures $W_{1}(S^A,S^B)=0$ and $W_{2}(S^A,S^B)=0$ in the $S^A-S^B$ plane for any given values of $p$, $\langle k\rangle$ and $\beta$ as shown in Fig.\ref{f2}.

When $\beta >\beta_c'$, we find that cures $W_{1}=0$ and $W_{2}=0$ have a tangent point with $S^A_c>0$ and $S^B_c>S^B_0=0$ (see panels $(a)-(c)$ of Fig.\ref{f2}). This indicates that the system undergoes a discontinuous percolation transition when $\beta >\beta_c'$. For $\beta<\beta_c'$, $S^B_0>0$, there exists two cases shown in panels $(d)-(f)$ and $(g)-(i)$ of Fig.\ref{f2}, respectively. For panels $(g)-(i)$, the tangent point of cures $W_{1}=0$ and $W_{2}=0$ appears with $S^A_c>0$ and $S^B_c>S^B_0>0$, which indicates the system also undergoes a discontinuous percolation transition for $\beta=0.2$. However, for $\beta=0.4$ ($(d)-(f)$ of Fig.\ref{f2}), the nontrivial cross point of cures $W_{1}=0$ and $W_{2}=0$ appears at $S^A_c=0$ and $S^B_c=S^B_0>0$. This means that the system undergoes a continuous percolation transition, when $\beta$ is larger than a certain value $\beta_c(<\beta_c')$.

In the following, we try to obtain the two tricritical points of the system as indicated in Fig.\ref{f2}. In general, we can keep $S^B$ constant in function $W_1$, and check the behaviours of the order parameter $S^A$. In this way, it is easy to know that the critical point $p_c$ must satisfy the derivative of equation $W_{1}(S^A,S^B)=0$ with respect to $S^A$, that is
\begin{equation}
2(1-\beta) \langle k\rangle p_c^2e^{-\langle k\rangle S^A_c}(1-e^{-\langle k\rangle S^A_c})+\beta \langle k\rangle p_c e^{-\langle k\rangle S^A_c} (1-e^{-\langle k\rangle S^B_c}) = 1. \label{w1d}
\end{equation}
It is obvious that this equation will hold for the value $(S^A_c,S^B_c)$. For the discontinuous percolation transition, we don't know the simple form of $(S^A_c,S_c^B)$, which can be obtained numerically as shown in Fig.\ref{f2}. So, we put our attention to the continuous percolation transition, for which $S_c^A=0$ and $S_c^B=S^B_0$. A simple calculation will tell us that $S^B_0=0$ does not make eq.(\ref{w1d}) true. Conclusion can be drawn that the continuous percolation transition can only be found when $\beta<\beta_c'$, i.e., $\beta_c'$ is one of the tricritical points.

As discussed earlier, when $\beta<\beta_{c}'$, the system does not always take a continuous percolation transition. This phenomenon is similar with the findings in refs.\cite{idn2,lim1,lim2}. As shown in these papers, this type of tricritical point also satisfies $d^2W_{1}(S^A,S^B)/d(S^{A})^2=0$. Note that at the tricritical point, the conditions of continuous and discontinuous percolation transitions are satisfied simultaneously. Hence, we have
\begin{equation}
\langle k\rangle (1-e^{-\langle k\rangle S^B_0})^{2}\beta_c^{2}+2\beta_c-2=0.
\end{equation}
That is
\begin{equation}
\beta_{c}=\frac{\sqrt{1+2\langle k\rangle(1-e^{-\langle k\rangle S^B_0})^{2}}-1}{\langle k\rangle(1-e^{-\langle k\rangle S^B_0})^{2}}, \label{betac}
\end{equation}
where $S^B_0$ can be obtained by eq.(\ref{sb0}). Above all, the system demonstrates a continuous percolation transition for $\beta_c<\beta<\beta_{c}'$, and discontinuous percolation transition for $\beta<\beta_c$ or $\beta>\beta_{c}'$.

In addition, we can also get the continuous percolation transition point from eq.(\ref{w1d}) by letting $S^A_c=0$ and $S^B_c=S^B_0$,
\begin{equation}
p_{c}^{II}=\frac{1}{\beta \langle k\rangle (1-e^{-\langle k\rangle S^B_0})}. \label{pc2}
\end{equation}
For discontinuous percolation transition, the critical point $p_{c}^{I}$ can be obtained numerically as shown in Fig.\ref{f2}.

Since $S^B_0$ decreases with the increase of $\beta$ as shown in Fig.\ref{f1}, there is a typical $\beta^*$ that minimizes the critical point $p_{c}^{II}$ (see eq.(\ref{pc2})), which corresponds to the optimal robustness of the system. The optimal solution $\beta^*$ can also be obtained numerically by eqs.(\ref{sb0}) and (\ref{pc2}), some simulation results will be shown later.

Furthermore, we can find that with the decreasing of average degree, $\beta_c$ increases and $\beta_c'$ decreases. As a result, the two tricritical points can merge together when the average degree is less than a typical value $\langle \tilde{k}\rangle$, i.e., the continuous percolation transition disappears when $\langle k\rangle$ less than $\langle\tilde{k}\rangle$. This typical value $\langle \tilde{k}\rangle$ can be easily found by letting $\beta_{c}=\beta_{c}'$.  Substituting $\beta_{c}\approx1-2.4554/\langle k\rangle$ and $S^B_0\approx1.2564/\langle k\rangle$ into eq.(\ref{pc2}), we can get the typical average degree $\langle\tilde{k}\rangle\approx 5.5533$.

\subsection*{Scale-free networks}
For scale-free networks, the degree distribution is $P(k)\sim k^{-\lambda}(k_{min} \leq k \leq k_{max})$, where $k_{min}$ and $k_{max}$ are the lower and upper bounds of the degree, respectively, and $\lambda$ is the power law exponent. The sizes of the giant components $S^{A}$ and $S^{B}$ can be solved numerically by using the theoretical framework developed in eqs. (\ref{Sa}) and (\ref{Sb}). Since the sizes of giant components $S^{A}$ and $S^{B}$ depend on the auxiliary parameters $R^A$ and $R^B$ directly, we can discuss the phase transition of the system by using the parameters $R^A$ and $R^B$. In order to locate the tricritical points $\beta_c$ and $\beta_{c}'$ for two coupled scale-free networks, we use the similar methods as the coupled random networks. We keep $R^B$ constant in eq. (\ref{Ra}), and check the behaviours of the order parameter $R^A$. At the critical point $p_c$, we have
\begin{equation}
p_c^2 (1-\beta) \{G^{A\prime}_{1}(1-R^{A}_{c})[1-G^{A}_{0}(1-R^{A}_{c})]+[1-G^{A}_{1}(1-R^{A}_{c})]G^{A\prime}_{0}(1-R^{A}_{c})\}+p_c \beta G^{A\prime}_{1}(1-R^{A}_{c})[1-G^{B}_{0}(1-R^{B}_{c})]=1. \label{sfw1d}
\end{equation}
For the continuous percolation transition, $R_c^A=0$ and $R_c^B=R^B_0$ with $R^B_0\neq0$. When $R^B_0=0$, eq. (\ref{sfw1d}) cannot hold any more, and we can conclude that $\beta_{c}'$, at which $R^B_{0}$ jumps to zero, is also one of the tricritical points. At this time, we can get the continuous percolation transition point from eq.(\ref{sfw1d})
\begin{equation}
p^{II}=\frac{1}{\beta \frac{\langle k(k-1)\rangle}{\langle k\rangle} [1-G^{B}_{0}(1-R^{B}_{0})]}. \label{sfpc2}
\end{equation}
Similar to the coupled random networks, $\beta_{c}'$  and $R_0^B$ can be solved numerically by letting $R_c^A=0$ in eq. (\ref{Rb}), therefore, we have
\begin{equation}
R_{0}^{B}=(1-\beta)[1-G^{B}_{1}(1-R^{B}_{0})][1-G^{B}_{0}(1-R^{B}_{0})]. \label{sfsb0}
\end{equation}

At the other tricritical point $\beta_{c}$, the conditions of continuous and discontinuous percolation transitions are satisfied simultaneously, i.e., $\beta_{c}$ makes the first and the second order derivative of eq. (\ref{Ra}) with respective $S^{A}$ hold at the percolation transition point $p_{c}$. Hence we have
\begin{equation}
[1-G^{B}_{0}(1-R^{B}_{c})]^{2}\frac{\langle k(k-1)(k-2) \rangle}{\langle k\rangle^{2}}\beta_{c}^{2}+2\beta_{c}-2=0.
\end{equation}
The critical point $\beta_{c}$ is
\begin{equation}
\beta_{c}=\frac{\sqrt{1+2[1-G^{B}_{0}(1-R^{B}_{c})]^{2}\frac{\langle k(k-1)(k-2) \rangle}{\langle k\rangle^{2}}}-1}{[1-G^{B}_{0}(1-R^{B}_{c})]^{2}\frac{\langle k(k-1)(k-2) \rangle}{\langle k\rangle^{2}}}. \label{sfbetac}
\end{equation}

By plugging the degree distribution for scale-free networks into the generating functions, we can get the theoretical values for the tricritical point $\beta_{c}$, the second order percolation points $p^{II}$, as well as the numerical solution for $\beta_{c}'$. Similar to random networks, we cannot get the analytical expressions for the first order percolation transition points, but they can be solved numerically by eq. (\ref{Ra}).

\subsection*{Simulation results and discussion}

\begin{figure}
\scalebox{0.6}[0.6]{\includegraphics{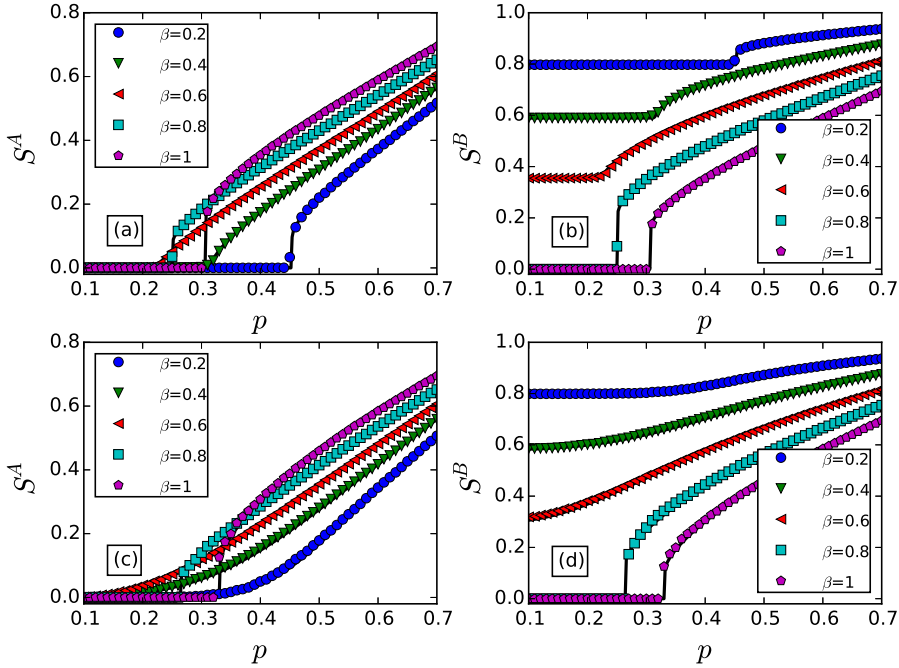}} \caption{ (Colour online) The sizes of the giant components $S^A$ and $S^B$ \emph{vs.} $p$. Panels (a) and (b) show the results for network $A$ and network $B$ in coupled random networks with $\langle k \rangle =8$, respectively. Panels (c) and (d) show the results for network $A$ and network $B$ in coupled scale-free networks with $k_{min}=4$, $k_{max}=316$ and $\lambda=2.7$, respectively. The solid lines show the theoretical predictions, and the symbols represent simulation results from $20$ time realizations on networks with $10^5$ nodes.} \label{f3}
\end{figure}

We firstly show how the giant component sizes $S^A$ and $S^B$ vary in dependence on the fraction of initial preserved nodes $p$ for both coupled random networks and coupled scale-free networks by simulation and theory in Fig.\ref{f3}. One can find that the analytical results are in agreement with the simulation results well. For the results of coupled random networks, one can find that the giant component size $S^A$ of network $A$ emergences abruptly when $p$ exceeds a threshold $p_{c}^{I}$ for $\beta=0.2$, $\beta=0.8$ and $\beta=1$. However, for $\beta=0.4$ and $\beta=0.6$, the giant component size $S^A$ of network $A$ increases continuously as $p$ exceeds a threshold $p_{c}^{II}$. The phenomena of network $B$ are similar, but a nonzero $S^B$ below the critical point for $\beta<\beta'_{c}$. For two coupled scale-free networks, the results are similar to the random networks, but different crtical points and tricritical points. As the scale-free networks we used in Fig. \ref{f3}, $\langle k(k-1)\rangle$ is divergence for a network with infinite size. Hence, according to eq.\ref{sfpc2}, the second order critical point $p_c^{II}\rightarrow0$.

\begin{figure}
\scalebox{0.6}[0.6]{\includegraphics{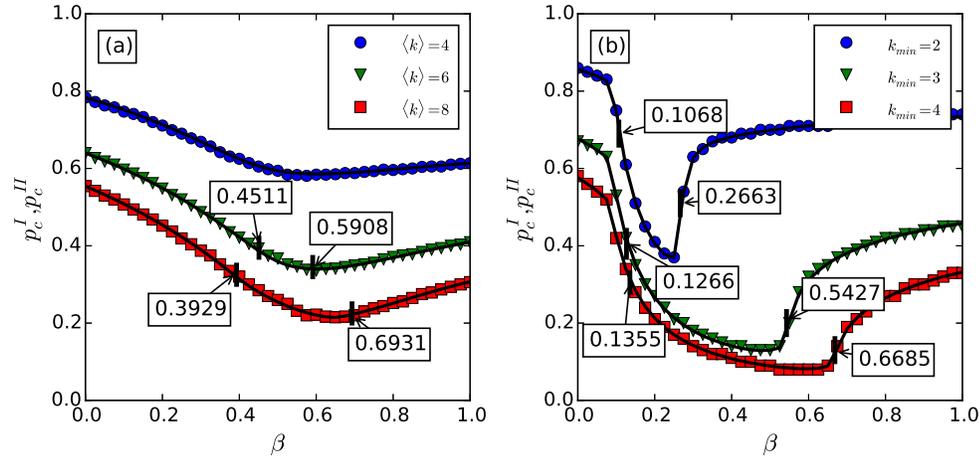}} \caption{ (Colour online) The critical point $p_c$ for different values of $\beta$. Panel (a) shows the results for coupled random networks with different average degree. For $\langle k\rangle=8$, the first tricritical point $\beta_{c}=0.3929$ and the second tricritical point $\beta_{c}'=0.6931$. For $\langle k\rangle=6$, $\beta_{c}=0.4511$ and $\beta_{c}'=0.5908$. For $\langle k\rangle=4$, the two tricritical points are merged together and the coupled networks always demonstrate discontinuous percolation transition. The theoretical prediction for the continuous percolation transition points $p_{c}^{II}$ are the results of eq.(\ref{pc2}) and the discontinuous percolation transition points $p_c^I$ are obtained as the way shown in Fig.\ref{f2}. Panel (b) shows the results for coupled scale-free networks with different lower bounds $k_{min}$ and the same upper bound $k_{min}=316$. For $k_{min}=2$, $\beta_{c}=0.1068$ and $\beta_{c}'=0.2633$. For $k_{min}=3$, $\beta_{c}=0.1266$ and $\beta_{c}'=0.5427$. For $k_{min}=4$, $\beta_{c}=0.1355$ and $\beta_{c}'=0.6685$. In both panels, the symbols represent simulation results from $20$ time realizations on networks with $10^5$ nodes, and the solid lines represent the theoretical predictions. } \label{f4}
\end{figure}

From Fig.\ref{f3}, we can also find that the threshold $p_{c}$ first decreases and then increases along with the increasing of $\beta$ for both coupled random networks and coupled scale-free networks, which can be further validated in Fig.\ref{f4}. Since the impact of initial removal is different for networks $A$ and $B$, the significance of the phenomenon is also slight different. For network $A$ suffered attack, its robustness can be optimized by arranging the ratio of inter-dependency links and inner-dependency links properly. For network $B$, the impacts of the initial node removal can be reduced by decreasing the fraction of inter-dependency links, however, more inner-dependency links will also reduce the stability of network $B$ itself. Note that all the second critical points of SF networks shown in Fig.\ref{f4} will be zero, when the network size tends to infinite.

The phase diagrams of the systems, including coupled random networks and coupled scale-free networks, are shown in Fig.\ref{f4} by both simulation and analysis. We use the simulation method developed by Parshani \emph{et al.} to estimate the discontinuous percolation transition points \cite{sdl1}. That is the number of iterative failures (NOI) sharply increases with approaching the critical point $p^{I}_{c}$. For the continuous transition, we calculate the point of maximum fluctuation for the size of the giant component to estimate the critical transition point \cite{idn8}. From Fig.\ref{f4}, one can find that the simulation and theoretical results are consistent well, and there is an optimal $\beta^*$ to maximize the system robustness for both coupled random networks and scale-free networks. This shows that a suitable arrangement of the dependency links will suppress the prorogation of failure within and among networks, simultaneously. Furthermore, this finding also indicates that the high density of dependency links could not always lead to a discontinuous percolation transition as the previous studies \cite{idn2,sdl1}. In addition, for coupled random networks, one can also find that the crossover of the two types of percolation transitions disappears as our theory prediction, when the average degree is below $\langle\tilde{k}\rangle\approx 5.5533$. For coupled scale-free networks, the crossover of the two types of percolation transitions can also disappear, the condition for which depends on the degree distributions of the coupled networks. Furthermore, critical exponents of a percolation system depend on its dimension \cite{percolation}. For random graphs and scale-free networks, they can be regarded as infinite dimensional systems, and their critical exponents are mean field and belong to the same universality class.

\section*{Conclusions}
In this paper we have studied the cascading failures in coupled networks with each node has a inner-dependency or inter-dependency link. Through simulation and theoretical study, we found that there exists an optimal value of $\beta^*$ leading to the most robust coupled networks for both random networks and scale-free networks, where $\beta$ is the fraction of the nodes have inter-dependency links.

More interestingly, we found that the high density of dependency links does not always lead to a discontinuous percolation transition as the previous studies. For random coupled networks, as long as the average degree of the network exceeds a typical $\langle\tilde{k}\rangle\approx5.5533$, the system will demonstrate a continuous percolation transition for $\beta_{c}<\beta<\beta_{c}'$, where the two tricritical points $\beta_{c}$ and $\beta_{c}'$ can be obtained exactly by our theoretical method. These results reveal that the number of dependency links is not the only factor that affects the robustness of the coupled networks, and a suitable arrangement of the dependency links will suppress the prorogation of failure within and among networks, simultaneously. We think that this nontrivial combined effect of the two types dependency links shown in this work will facilitate the design of resilient infrastructures.

\section*{Acknowledgements}

This work is funded by: The National Natural Science Foundation of China (Grant Nos.: 11305042, 61503355, 61403114, 11275186). RRL and CXJ acknowledge the support of the research start-up fund of Hangzhou Normal University (Grant Nos.: 2011QDL29, 2011QDL31). ML is also supported by the Fundamental Research Fund for the Central Universities. CXJ is also supported by the Zhejiang Provincial Natural Science Foundation of China under Grant No. LQ14F030009.

\section*{Author contributions statement}
RRL, ML, CXJ, and BHW conceived and designed the research. RRL and CXJ carried out the numerical simulations. RRL and ML developed the theory and wrote the manuscript.

\section*{Additional Information}
The authors declare no competing financial interests.


\begin{thebibliography}{99}
\bibitem{nw3} Albert, R., Jeong, H. \& Barab\'{a}si, A.-L. Error and attack tolerance of complex networks. {\it Nature} {\bf 406}, 378-382 (2000).
\bibitem{nw8} Newman, M. E. J. {\it Networks: An Introduction} (Oxford University Press, Oxford, 2010)
\bibitem{nw9} Cohen, R. \& Havlin, S. Complex Networks: Structure, Robustness and Function (Cambridge University Press, 2010).
\bibitem{idn1} Buldyrev, S. V., Parshani, R., Paul, G. \& Stanley, H. E. Catastrophic failures in interdependent networks. {\it Nature} {\bf 464}, 1025-1028 (2010).
\bibitem{vespignani2010} Vespignani, A. Complex networks: The fragility of interdependency. {\it Nature} {\bf 464}, 984-985 (2010).
\bibitem{Gao2011a} Gao, J., Buldyrev, S. V., Stanley, H. E. \& Havlin, S. Networks formed from interdependent networks. {\it Nature Phys.} {\bf 8}, 40-48 (2011).
\bibitem{Gao2011b} Gao, J., Buldyrev, S. V., Havlin, S. \& Stanley, H. E. Robustness of a network of networks. {\it Phys. Rev. Lett.} {\bf 107}, 195701 (2011).
\bibitem{non1} Havlin, S., Stanley, H. E., Bashan, A., Gao, J., \& Kenett, D. Y. Percolation of interdependent network of networks. {\it Chaos, Solitons \& Fractals} {\bf 72}, 4-19(2015).
\bibitem{non2} Bianconi, G., \& Dorogovtsev, S. N. Multiple percolation transitions in a configuration model of a network of networks. {\it Phys. Rev. E} {\bf 89}, 062814 (2014).
\bibitem{idn2} Parshani, R., Buldyrev, S. \& Havlin, S. Interdependent Networks: Reducing the coupling strength leads to a change from a first to second order percolation transition. {\it Phys. Rev. Lett.} {\bf 105}, 048701 (2010).
\bibitem{idn3} Parshani, R., Rozenblat, C., Ietri, D., Ducruet, C. \& Havlin, S. Inter-similarity between coupled networks. {\it Europhys. Lett} {\bf 92}, 2470-2484 (2010)
\bibitem{idn4} Hu, Y. \emph{et al}. Percolation of interdependent networks with intersimilarity. {\it Phys. Rev. E} {\bf 88}, 052805 (2013).
\bibitem{idn5} Shao, J., Buldyrev, S. V., Havlin, S. \& Stanley H. E. Cascade of failures in coupled network systems with multiple support-dependence relations. {\it Phys. Rev. E} {\bf 83}, 036116 (2011).
\bibitem{idn6} Huang, X., Gao, J., Buldyrev, S. V., Havlin, S. \& Stanley, H. E. Robustness of interdependent networks under targeted attack. {\it Phys. Rev. E} {\bf 83}, 065101(R) (2011).
\bibitem{loc1} Berezin, Y., Bashan, A., Danziger, M. M., Li, D. \& Havlin, S. Localized attacks on spatially embedded networks with dependencies. {\it Sci. Rep.} {\bf 5}, 08934 (2015)
\bibitem{loc2} Shao, S., Huang, Q., Stanley, H. E. \& Havlin, S. Percolation of localized attack on complex networks. {\it New J. Phys.} {\bf 17}, 023049 (2015)
\bibitem{idn8} Zhou, D., Stanley, H. E., D'Agostino, G., \& Scala, A., Simultaneous first- and second-order percolation transitions in interdependent networks. {\it Phys. Rev. E} {\bf 86}, 066103 (2012).
\bibitem{Valdez} Valdez, L. D., Macri, P. A., Stanley, H. E., \& Braunstein, L. A., Triple point in correlated interdependent networks. {\it Phys. Rev. E} {\bf 88}, 050803(R) (2013).
\bibitem{assor} Min, B., Yi, S. D., Lee, K.-M., \& Goh, K.-I. Network robustness of multiplex networks with interlayer degree correlations. {\it Phys. Rev. E} {\bf 89}, 042811 (2014).
\bibitem{cluster1} Shao, S., Huang, X., Stanley, H. E., \& Havlin, S. Robustnessof a partially interdependent network formed of clustered networks. {\it Phys. Rev. E} {\bf 89}, 032812 (2014).
\bibitem{cluster2} Huang, X. \emph{et al}. The robustness of interdependent clustered networks. {\it Europhys. Lett.} {\bf 101}, 18002 (2013)
\bibitem{de1} Emmerich, T., Bunde, A., \& Havlin, S. Structural and functional properties of spatially embedded scale-free networks. {\it Phys. Rev. E} {\bf 89}, 062806 (2014).
\bibitem{de2} Yuan, X., Shao, S., Stanley, H. E., \& Havlin, S. How breadth of degree distribution influences network robustness: comparing localized and random attacks. {\it Phys. Rev. E} {\bf 92}, 032122 (2015).
\bibitem{idn9} Li, W., Bashan, A., Buldyrev, S. V., Stanley, H. E. \& Havlin, S. Cascading failures in interdependent lattice networks: the critical role of the length of dependency links. {\it Phys. Rev. Lett.} {\bf 108}, 228702 (2012).
\bibitem{Bashan2013} Bashan, A., Berezin, Y., Buldyrev, S. V. \& Havlin, S. The extreme vulnerability of interdependent spatially embedded networks. {\it Nature Phys.} {\bf 9}, 667-672 (2013).
\bibitem{Shekhtman} Shekhtman, L. M., Berezin, Y., Danziger, M. M. \& Havlin, S. Robustness of a network formed of spatially embedded networks. {\it Phys. Rev. E} {\bf 90}, 012809 (2014).
\bibitem{Danziger} Danziger, M. M., Bashan, A., Berezin, Y., Havlin, S. Percolation and cascade dynamics of spatial networks with partial dependency.  {\it J. Complex Networks} {\bf 2}, 460-474 (2014).
\bibitem{sdl1} Parshani, R., Buldyrev, S. V. \& Havlin, S. Critical effect of dependency groups on the function of networks. {\it Proc. Natl. Acad. Sci. U.S.A.} {\bf 108}, 1007-1010 (2011).
\bibitem{lim1} Li, M., Liu, R.-R., Jia, C.-X., \& Wang, B.-H. Critical effects of overlapping of connectivity and dependence links on percolation of networks. {\it New J. Phys.} {\bf 15}, 093013 (2013).
\bibitem{sdl2} Bashan, A., Parshani, R. \& Havlin, S. Percolation in networks composed of connectivity and dependency links. {\it Phys. Rev. E} {\bf 83}, 051127 (2011).
\bibitem{sdl3} Bashan, A. \& Havlin, S. The combined effect of connectivity and dependency links on percolation of networks. {\it Journal of Statistical Physics.} {\bf 145}, 686-695 (2011).
\bibitem{lim2} Li, M., Liu, R.-R., Jia, C.-X., \& Wang B.-H. Cascading failures on networks with asymmetric dependence. {\it Europhys. Lett.} {\bf 108}, 56002 (2013).
\bibitem{lim3} Wang, H., Li, M., Deng, L., \& Wang, B.-H.. Percolation on networks with conditional dependence group. {\it PLoS ONE} {\bf 10}, e0126674 (2015).
\bibitem{rvn} Boccaletti, S. \emph{et al}. The structure and dynamics of multilayer networks. {\it Phys. Rep.} {\bf 544}, 1 (2014).
\bibitem{fs1} Son, S.-W., Grassberger, P., \& Paczuski, M. Percolation transition are not always sharpened by making networks interdependent, {\it Phys. Rev. Lett.} {\bf 107}, 195702 (2011).
\bibitem{fs2} Son, S.-W., Bizhani, G., Christensen, C., Grassberger, P. \& Paczuski, M. Percolation theory on interdependent networks based on epidemic spreading. {\it Europhys. Lett.} {\bf 97}, 16006 (2012).
\bibitem{nm} Newman, M. E. J., Strogatz, S. H., \& Watts, D. J. Random graphs with arbitrary degree distributions and their applications. {\it Phys. Rev. E} {\bf 64}, 026118 (2001).
\bibitem{tmgt} Bollob\'{a}s, B. {\it Random Graphs} (Academic Press, London, 1985).
\bibitem{percolation} Stanley, H. E. {\it Introduction to Phase Transitions and Critical Phenomena} (Oxford University Press, 1971).


\end{thebibliography}
\end{document}